\documentclass[a4paper,11pt]{article}

\usepackage{jheppub}
\usepackage[T1]{fontenc}
\usepackage{hyperref}
\usepackage{graphicx}
\usepackage{amsmath}
\usepackage{epsfig}
\usepackage{subfig}
\usepackage{xcolor}
\usepackage{natbib}
\usepackage{multirow}

\def\bds{\boldsymbol}

\title{Next-to-leading-order study of $J/\psi$ angular distributions in $e^{+}e^{-} \to J/\psi+\eta_c,\chi_{cJ}$ at $\sqrt{s} \approx 10.6$ GeV}
\author{Zhan Sun}
\affiliation{Department of Physics, Guizhou Minzu University, Guiyang 550025, People's Republic of China.}
\emailAdd{zhansun@cqu.edu.cn}

\abstract{In this paper, we present a detailed next-to-leading-order (NLO) study of $J/\psi$ angular distributions in $e^{+}e^{-} \to J/\psi+\eta_c,\chi_{cJ}$ ($J=0,1,2$) within the nonrelativistic QCD factorization (NRQCD). The numerical NLO expressions for total and differential cross sections, i.e., $\frac{d\sigma}{d\cos\theta}=A+B\cos^2\theta$, are both derived. With the inclusion of the newly-calculated QCD corrections to $A$ and $B$, the $\alpha_{\theta}(= B/A)$ parameters in $J/\psi+\chi_{c0}$ and $J/\psi+\chi_{c1}$ are moderately enhanced, while the magnitude of ${\alpha_\theta}_{J/\psi+\chi_{c2}}$ is significantly reduced; regarding the production of $J/\psi+\eta_c$, the $\alpha_\theta$ value remains unchanged. By comparing with experiment, we find the predicted ${\alpha_\theta}_{J/\psi+\eta_c}$ is in good agreement with the $\textrm{B}\scriptsize{\textrm{ELLE}}$ measurement; however, ${\alpha_\theta}_{J/\psi+\chi_{c0}}$ is still totally incompatible with the experimental result, and this discrepancy seems to hardly be cured by proper choices of the charm-quark mass, the renormalization scale, and the NRQCD matrix elements.}


\begin{document}

\maketitle

\bibliographystyle{JHEP}

\section{Introduction}\label{intro}

In the past twenty years, the $\textrm{B}\scriptsize{\textrm{ELLE}}$ and $\textrm{B}\scriptsize{\textrm{ABAR}}$ Collaborations have independently measured the total cross sections of $e^{+}e^{-} \to J/\psi+\eta_c,\chi_{c0}$ at $\sqrt{s} \approx 10.6$ GeV \cite{Abe:2002rb,Abe:2004ww,Aubert:2005tj}, which significantly overshoot the results \cite{Braaten:2002fi,Liu:2004ga,Liu:2002wq,Hagiwara:2003cw} calculated at leading order (LO) in $\alpha_s$ using the nonrelativistic QCD (NRQCD) factorization \cite{NRQCD}. As a breakthrough of the theoretical attempts \cite{Ma:2004qf,Bondar:2004sv,Braguta:2005kr,Bodwin:2006dm,Zhang:2005cha,Choi:2007ze,Bodwin:2007ga,Gong:2007db,Zhang:2008gp,Braguta:2008tg,Sun:2009zk,Brambilla:2010cs,Wang:2011qg,Dong:2011fb,Dong:2012xx,Wang:2013vn,Xi-Huai:2014iaa,Brambilla:2014jmp,Bodwin:2014dqa,Sun:2018rgx,Jiang:2018wmv,Lansberg:2019adr,Zeng:2021hwt,Tao:2019rwy,Berezhnoy:2021tqb} to explain this inconsistency, the next-to-leading-order (NLO) QCD  corrections \cite{Zhang:2005cha,Gong:2007db,Zhang:2008gp} can provide considerable and positive contributions, largely alleviating the discrepancies between theory and experiment.

Besides the total cross sections, based on a data sample of $140~\textrm{fb}^{-1}$, $\textrm{B}\scriptsize{\textrm{ELLE}}$ has also measured the $J/\psi$ angular distributions in $e^{+}e^{-} \to J/\psi+\eta_c,\chi_{c0}$ \cite{Abe:2004ww}, i.e., the ${\alpha_\theta}$ parameter in $\frac{d\sigma}{d\cos\theta}=A+B\cos^2\theta=A(1+{\alpha_\theta}\cos^2\theta)$. By simultaneously fitting the production- and helicity-angle distributions, the measured ${\alpha_\theta}$ reads
\begin{eqnarray}
{\alpha_\theta}(e^{+}e^{-} \to J/\psi+\eta_c)&=&0.93^{+0.57}_{-0.47}, \nonumber \\
{\alpha_\theta}(e^{+}e^{-} \to J/\psi+\chi_{c0})&=&-1.01^{+0.38}_{-0.33}.
\label{eq: alpha measurements}
\end{eqnarray}
On the theoretical side, the existing studies of the differential cross section ($\frac{d\sigma}{d\cos\theta}$) are only accurate to the LO level in $\alpha_s$ \cite{Braaten:2002fi,Liu:2004ga}. For $J/\psi+\eta_c$, the LO calculations using NRQCD give a prediction ${\alpha_\theta}=1$, which is consistent with the measured value in equation (\ref{eq: alpha measurements}); however, the NRQCD-based LO predictions of ${\alpha_\theta}_{(J/\psi+\chi_{c0})} \simeq 0.25$ are fundamentally different from the above $\textrm{B}\scriptsize{\textrm{ELLE}}$ measurement. In order to fill the huge gap between theory and experimental result, spurred by the significant impacts of the QCD corrections on total cross section, it is urgent to carry out a NLO analysis to the differential cross section. Moreover, whether the inclusion of the uncalculated QCD corrections would spoil the existing coincidence of ${\alpha_\theta}_{J/\psi+\eta_c}$ with experiment need also to be verified.

At this stage, $\textrm{B}\scriptsize{\textrm{ELLE}}$ and $\textrm{B}\scriptsize{\textrm{ABAR}}$ have not observed any evident event of $e^{+}e^{-} \to J/\psi+\chi_{c1,2}$, owing to the relatively small production rates of the two channels. Fortunately, the recently-commissioning Super-$B$ factories with a high luminosity designed to reach up to about $50~\textrm{ab}^{-1}$ by 2022 would bring great opportunities to fulfill the observations, which can aid in further understanding the double-charmonium productions in $e^{+}e^{-}$ annihilation. Taken together, in this paper we will for the first time perform a comprehensive NLO study of the differential cross sections in $e^{+}e^{-} \to J/\psi+\eta_c,\chi_{cJ}$ with $J=0,1,2$.

Note that, in the context of NRQCD factorization, due to the inadequate knowledge about the heavy-quarkonium production mechanism, the calculated $\sigma_{e^{+}e^{-} \to J/\psi+\eta_c,\chi_{cJ}}$ suffer severely from the indeterminacy inherent to the nonperturbative NRQCD long distance matrix elements (LDMEs), which would then significantly weaken the predictive power of NRQCD. On the contrary, as a result of the cancellation of the dependences of $A$ and $B$ on LDME, the theoretical result of ${\alpha_\theta}(=B/A)$ dose NOT involve this kind of ambiguity. In this sense, the ${\alpha_\theta}$ parameter is expected to be a more ideal laboratory than total cross section for the study of exclusive double-charmonium productions in $e^{+}e^{-}$ annihilation. Considering the large uncertainty of ${\alpha_\theta}_{J/\psi+\eta_c,\chi_{c0}}$ in equation (\ref{eq: alpha measurements}) and the lack of measured ${\alpha_\theta}_{J/\psi+\chi_{c1,2}}$, we strongly suggest the Super-$B$ factories to reperform with better precision the measurements of $\alpha_\theta$, and our state-of-the-art calculation results would pave the way for the future comparisons.

The rest of the paper is organized as follows: Section \ref{cal} is an outline of the calculation formalism. Then, the phenomenological results and discussions are presented in Section \ref{results}. Section \ref{sum} is reserved as a summary.

\section{Calculation formalism}\label{cal}

Following the NRQCD factorization, the differential cross sections of $e^{+}(p_1)+e^{-}(p_2) \to J/\psi(p_3)+\eta_c,\chi_{cJ}(p_4)$ can be generally written as
\begin{eqnarray}
d\sigma=d\hat{\sigma}_{e^{+}e^{-} \to c\bar{c}[n_1]+c\bar{c}[n_2]}\langle \mathcal{O}^{J/\psi}(n_1)\rangle \langle \mathcal{O}^{\eta_c(\chi_{cJ})}(n_2)\rangle,
\end{eqnarray}
where $d\hat{\sigma}_{e^{+}e^{-} \to c\bar{c}[n_1]+c\bar{c}[n_2]}$ is the perturbative calculable short distance coefficients, denoting the production of a configuration of $c\bar{c}[n_1]$ intermediate state associated with $c\bar{c}[n_2]$. By neglecting the color-octet contributions, which are discovered to be trivial for the production of $e^{+}+e^{-} \to J/\psi+\eta_c(\chi_{cJ})$ \cite{Braaten:2002fi}, $n_1=^3S_1^{1}$ and $n_2=^1S_0^{1}(^3P_J^{1})$. The universal nonperturbative LDMEs $\langle \mathcal{O}^{J/\psi}(n_{1})\rangle$ and $\langle \mathcal{O}^{\eta_c(\chi_{cJ})}(n_{2})\rangle$ stand for the probabilities of $c\bar{c}[n_{1}]$ and $c\bar{c}[n_{2}]$ into $J/\psi$ and $\eta_c(\chi_{cJ})$, respectively.

$d\hat{\sigma}_{e^{+}e^{-} \to c\bar{c}[n_1]+c\bar{c}[n_2]}$ can be further expressed as
\begin{eqnarray}
d\hat{\sigma}_{e^{+}e^{-} \to c\bar{c}[n_1]+c\bar{c}[n_2]}=|\mathcal{M}|^2 d\Pi_{2}=L_{\mu\nu}H^{\mu\nu} d\Pi_{2},
\end{eqnarray}
where $L_{\mu\nu}$ and $H^{\mu\nu}$ are the leptonic and hadronic tensors, respectively, and $d\Pi_{2}$ is the standard two-bodies phase space.

\subsection{Leptonic current}

$L_{\mu\nu}$ can directly be calculated and obtained as
\begin{eqnarray}
L_{\mu\nu}&=&4 \pi \alpha \textrm{Tr}[p\!\!\!\slash_{1} \gamma_{\mu} p\!\!\!\slash_{2} \gamma_{\nu}] \nonumber \\
&=&4 \pi \alpha s \left( -2g_{\mu\nu}+\frac{4 {p_1}_{\mu}q_{\nu}+4 {p_1}_{\nu}q_{\mu}-8{p_1}_{\mu}{p_1}_{\nu}}{s} \right)\nonumber \\
&=&4 \pi \alpha s l_{\mu\nu},
\label{eq1}
\end{eqnarray}
where $q=p_1+p_2$ and $s=(p_1+p_2)^2$. {\it{Note that}}, in calculating total cross section, the standard $l_{\mu\nu}$ in equation (\ref{eq1}) can be replaced equivalently with $-\frac{8}{3}g^{\mu\nu}$ \cite{Gong:2009ng}, which, however, is no longer applicable for the case of differential cross section. Therefore, in order to evaluate $\frac{d\sigma}{d\cos\theta}$, we will employ a new $l_{\mu\nu}$ obtained in our recent paper \cite{Zhang:2017dia}; the following is a brief description of its derivation.  

By integrating $H^{\mu\nu}$ over all the final states other than $J/\psi$, we obtain the hadronic tensor $W_{h}^{\mu\nu}(p_3,q)$, which is dependent only on $p_3$ and $q$. Subsequently we decompose $W_{h}^{\mu\nu}(p_3,q)$ as a linear combination of the tensors constituted of $-g^{\mu\nu}$, $p_3$, and $q$ as
\begin{eqnarray}
W_{h}^{\mu\nu}(p_3,q)=W_1\left(-g^{\mu\nu}+\frac{q^{\mu}q^{\nu}}{s}\right)+W_2\left[{\left(p_3-\frac{p_3 \cdot q}{s}q\right)}^{\mu}{\left(p_3-\frac{p_3 \cdot q}{s}q\right)}^{\nu}\right],
\end{eqnarray}
which satisfies the following relation
\begin{eqnarray}
q_{\mu}W_{h}^{\mu\nu}=0.
\end{eqnarray}
With some calculations, the contraction of $W_{h}^{\mu\nu}$ with the $l_{\mu\nu}$ in equation (\ref{eq1}) says
\begin{eqnarray}
l_{\mu\nu}W_{h}^{\mu\nu}=4W_1+2|\bds{p}_3|^2(1-\cos^2\theta)W_2,
\label{eq2}
\end{eqnarray}
where $\bds{p}_3$ is the three momenta of $J/\psi$, and $\theta$ is the angle between $\bds{p}_3$ and the spatial momentum of $e^{-}$ (or $e^{+}$) in the $e^{+}e^{-}$ center-of-mass frame.

It is easy to verify that, with
\begin{eqnarray}
l_{\mu\nu}=\mathcal{A}_1\left(-g_{\mu\nu}+\frac{q_{\mu}q_{\nu}}{s}\right)+\mathcal{A}_2\left[\frac{{\left(p_3-\frac{p_3 \cdot q}{s}q\right)}_{\mu}{\left(p_3-\frac{p_3 \cdot q}{s}q\right)}_{\nu}}{|{\bds{p}_3}|^2}\right],
\end{eqnarray}
where
\begin{eqnarray}
\mathcal{A}_1&=&1+\cos^{2}\theta, \nonumber \\
\mathcal{A}_2&=&1-3\cos^{2}\theta,
\label{A1A2}
\end{eqnarray}
one can reproduce the results in equation (\ref{eq2}). Further employing the current conservation will finally yield
\begin{eqnarray}
l_{\mu\nu}=\mathcal{A}_1(-g_{\mu\nu})+\mathcal{A}_2\left(\frac{{p_3}_{\mu}{p_3}_{\nu}}{|{\bds{p}_3}|^2}\right).
\label{eq3}
\end{eqnarray}
With the integration over $\cos\theta$, the term involving $A_2$ in equation (\ref{eq3}) vanishes, and the first term on the right-hand side will reduce to $-\frac{8}{3}g^{\mu\nu}$. Comparing the two leptonic tensors in equations (\ref{eq1}) and (\ref{eq3}), one can easily find the newly-derived $l_{\mu\nu}$ is related only to the hadronic-process momentum ($p_3$), which, by the absence of $p_1$ and $p_2$, would greatly reduce the complication in computing the differential cross sections, especially at the NLO accuracy.

\subsection{Cross sections}

Using the leptonic tensor in equation (\ref{eq3}), the differential cross sections of $e^{+}e^{-} \to J/\psi+\eta_c,\chi_{cJ}$ can be written in the following form
\begin{eqnarray}
\frac{d\sigma}{d\cos\theta}=A+B\cos^2\theta=\kappa\left(\mathcal{C}_1 \mathcal{A}_1+\mathcal{C}_2 \mathcal{A}_2\right),
\label{eq4}
\end{eqnarray}
and then we have
\begin{eqnarray}
A&=&\kappa(\mathcal{C}_1+\mathcal{C}_2), \nonumber \\
B&=&\kappa(\mathcal{C}_1-3\mathcal{C}_2),  \nonumber \\
\sigma &=& \frac{8}{3}\kappa \mathcal{C}_1.
\label{eqAB}
\end{eqnarray}
The universal factor $\kappa$ reads
\begin{eqnarray}
\kappa_{J/\psi+\eta_c} &=& \frac{2 \pi \alpha^2 \alpha_s^2 |R_{1S}(0)|^2 |R_{1S}(0)|^2 \sqrt{s^2-16 m_c^2 s}}{81 m_c^2 s^{2}}, \nonumber \\
\kappa_{J/\psi+\chi_{cJ}} &=& \frac{2 \pi \alpha^2 \alpha_s^2 |R_{1S}(0)|^2 |R^{'}_{1P}(0)|^2 \sqrt{s^2-16 m_c^2 s}}{27 m_c^2 s^{2}}.
\end{eqnarray}
$|R_{1S}(0)|$ and $|R^{'}_{1P}(0)|$ are the wave functions at the origin, which can be related to the NRQCD LDMEs by the following formulae:
\begin{eqnarray}
\langle \mathcal O^{J/\psi}(^3S_1^{1}) \rangle & \simeq & 3 \langle \mathcal O^{\eta_c}(^1S_0^{1}) \rangle = \frac{9}{2\pi}|R_{1S}(0)|^2, \nonumber  \\
\langle \mathcal O^{\chi_{cJ}}(^3P_J^{1}) \rangle  &=& (2J+1)\frac{3}{4\pi}|R^{'}_{1P}(0)|^2.
\end{eqnarray}

\subsubsection{LO}

\begin{figure}[!h]
\begin{center}
\hspace{0cm}\includegraphics[width=0.65\textwidth]{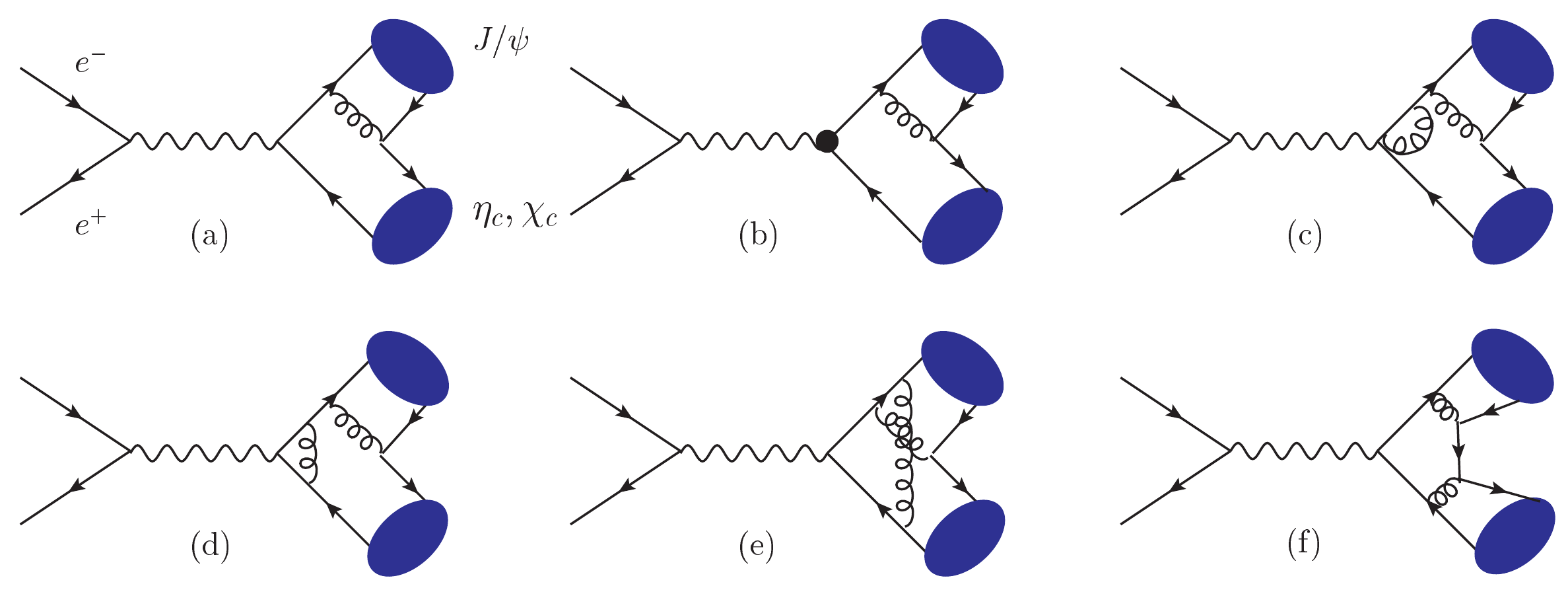}
\caption{\label{fig:Feynman Diagrams}
Typical LO and NLO Feynman diagrams for $e^{+}e^{-} \to J/\psi+\eta_c,\chi_{cJ}$.}
\end{center}
\end{figure}

According to the LO Feynman diagrams for $e^{+}e^{-} \to J/\psi+\eta_c,\chi_{cJ}$, which are free of divergence and which are representatively shown in figure \ref{fig:Feynman Diagrams}(a), we figure straightforwardly out the coefficients $\mathcal{C}_1$ and $\mathcal{C}_2$ through LO order in $\alpha_s$ for various processes,
\begin{itemize}
\item[(i)]
for $J/\psi+\eta_c$,
\begin{eqnarray}
\mathcal{C}_1&=&-\frac{128 r^3(4r-1)}{m_c^4}, \nonumber \\
\mathcal{C}_2&=&0,
\label{eq5}
\end{eqnarray}
\item[(ii)]
for $J/\psi+\chi_{cJ}$,
\begin{eqnarray}
\mathcal{C}^{J=0}_1&=&\frac{16 r^2 (144 r^4+152 r^3-428 r^2+128 r+1)}{3 m_c^6}, \nonumber \\
\mathcal{C}^{J=0}_2&=&\frac{16 r^2 (-12 r^2+10 r+1)^2}{3 m_c^6}, \nonumber \\
\mathcal{C}^{J=1}_1&=&\frac{128 r^3 (18 r^3+13 r^2-12 r+2)}{m_c^6}, \nonumber \\
\mathcal{C}^{J=1}_2&=&\frac{256 r^4 (1-3r)^2}{m_c^6}, \nonumber \\
\mathcal{C}^{J=2}_1&=&\frac{32 r^2 (360 r^4+308 r^3-188 r^2+20 r+1)}{3 m_c^6}, \nonumber \\
\mathcal{C}^{J=2}_2&=&\frac{32 r^2 (360 r^4-96 r^3+4 r^2-4 r+1)}{3 m_c^6},
\label{eq6}
\end{eqnarray}
\end{itemize}
where the dimensionless variable $r$ is defined as
\begin{eqnarray}
r & \equiv & \frac{4 m_c^2}{s}.
\end{eqnarray}
Following the relations in equation (\ref{eqAB}), one can directly obtain the analytical expressions of the LO-level $A$, $B$ and $\sigma$ concerning $e^{+}e^{-} \to J/\psi+\eta_c,\chi_{cJ}$, which can be proved to agree with the results in ref. \cite{Liu:2004ga}.

\subsubsection{NLO}

Due to the absence of the real-correction processes in NLO, we need only to calculate the virtual corrections, which include 60 one-loop diagrams and 20 counter-term diagrams, as illustrated in figures \ref{fig:Feynman Diagrams}(b)-\ref{fig:Feynman Diagrams}(f). We utilize the dimensional regularization with $D=4-2\epsilon$ to isolate the ultraviolet (UV) and infrared (IR) divergences. The on-mass-shell (OS) scheme is employed to set the renormalization constants for the $c$-quark mass ($Z_m$) and heavy-quark filed ($Z_2$); the minimal-subtraction ($\overline{MS}$) scheme is adopted for the QCD-gauge coupling ($Z_g$) and the gluon filed $Z_3$. The renormalization constants are taken as
\begin{eqnarray}
\delta Z_{m}^{OS}&=& -3 C_{F} \frac{\alpha_s N_{\epsilon}}{4\pi}\left[\frac{1}{\epsilon_{\textrm{UV}}}-\gamma_{E}+\textrm{ln}\frac{4 \pi \mu_r^2}{m_c^2}+\frac{4}{3}\right], \nonumber \\
\delta Z_{2}^{OS}&=& - C_{F} \frac{\alpha_s N_{\epsilon}}{4\pi}\left[\frac{1}{\epsilon_{\textrm{UV}}}+\frac{2}{\epsilon_{\textrm{IR}}}-3 \gamma_{E}+3 \textrm{ln}\frac{4 \pi \mu_r^2}{m_c^2}+4\right], \nonumber \\
\delta Z_{3}^{\overline{MS}}&=& \frac{\alpha_s N_{\epsilon}}{4\pi}(\beta_{0}-2 C_{A})\left[\frac{1}{\epsilon_{\textrm{UV}}}-\gamma_E+\textrm{ln}(4\pi)\right], \nonumber \\
\delta Z_{g}^{\overline{MS}}&=& -\frac{\beta_{0}}{2}\frac{\alpha_s N_{\epsilon}}{4\pi}\left[\frac{1} {\epsilon_{\textrm{UV}}}-\gamma_{E}+\textrm{ln}(4\pi)\right], \label{CT}
\end{eqnarray}
where $\gamma_E$ is the Euler's constant, $N_{\epsilon}= \Gamma[1-\epsilon] /({4\pi\mu_r^2}/{(4m_c^2)})^{\epsilon}$ is an overall factor in our calculation, and $\beta_{0}=\frac{11}{3}C_A-\frac{4}{3}T_Fn_f$ is the one-loop coefficient of the $\beta$ function. $n_f(=n_{L}+n_{H})$ represents the number of the active-quark flavors; $n_{L}$ and $n_{H}$ denote the number of the light- and heavy-quark flavors, respectively. At $\sqrt{s}=10.6$ GeV, $n_L=3$ and $n_H=1$. In ${\rm SU}(3)$, the color factors are given by $T_F=\frac{1}{2}$, $C_F=\frac{4}{3}$, and $C_A=3$.

By taking into account the QCD corrections, we acquire the NLO-level $\mathcal{C}_{1}$ and $\mathcal{C}_{2}$, and then $A$ and $B$ of NLO in $\alpha_s$ using the relations in equation (\ref{eqAB}). Since the fully analytical expressions of the NLO results are lengthy, we in the following just provide the numerical expressions, among which the choice of charm-quark mass $m_c=1.5 (\pm 0.1)$ GeV that is often used in calculating the charmonium-involved processes is chosen. 

The NLO expressions of $A$, $B$, or $\sigma$ can be generally written as
\begin{eqnarray}
(A,B,\textrm{or}~\sigma)^{\textrm{NLO}}=(A,B,\textrm{or}~\sigma)^{\textrm{LO}} \times \left[ 1+\frac{\alpha_s}{\pi} \left(\frac{1}{2}\beta_{0}\textrm{ln}\frac{\mu_r^2}{4m_c^2}+c_L n_{L}+c_H n_{H}+c \right) \right],
\label{NLOexp}
\end{eqnarray}
where the dimensionless coefficients $c_{L}$, $c_{H}$, and $c$ for various processes are summarized in tables \ref{tab: coefetac}, \ref{tab: coefxc0}, \ref{tab: coefxc1}, and \ref{tab: coefxc2}.

\begin{table*}[htb]
\begin{center}
\caption{The coefficients $c_{L}$, $c_{H}$, and $c$ for $e^{+}e^{-} \to J/\psi+\eta_c$ at $\sqrt{s}=10.6$ GeV. The unit of $m_c$ is GeV.}
\label{tab: coefetac}
\begin{tabular}{|c|ccc|ccc|ccc|cc}
\hline
$~$ & \multicolumn{3}{c|}{$A,B,\sigma$}\\ \hline
$m_c$ & $c_{L}$ & $c_{H}$ & $c$\\
$1.4$ & $-0.1302$ & $-0.3026$ & $13.066$\\
$1.5$ & $-0.1762$ & $-0.3782$ & $12.728$\\
$1.6$ & $-0.2192$ & $-0.4538$ & $12.436$\\
\hline
\end{tabular}
\end{center}
\end{table*}

\begin{table*}[htb]
\begin{center}
\caption{The coefficients $c_{L}$, $c_{H}$, and $c$ for $e^{+}e^{-} \to J/\psi+\chi_{c0}$ at $\sqrt{s}=10.6$ GeV. The unit of $m_c$ is GeV.}
\label{tab: coefxc0}
\begin{tabular}{|c|ccc|ccc|ccc|cc}
\hline
$~$ & \multicolumn{3}{c|}{$A$} & \multicolumn{3}{c|}{$B$} & \multicolumn{3}{c|}{$\sigma$}\\ \hline
$m_c$ & $c_{L}$ & $c_{H}$ & $c$ & $c_{L}$ & $c_{H}$ & $c$ & $c_{L}$ & $c_{H}$ & $c$\\
$1.4$ & $-0.2302$ & $-0.4617$ & $8.0679$ & $-0.0305$ & $-0.1440$ & $9.7089$ & $-0.2147$ & $-0.4372$ & $8.1947$\\
$1.5$ & $-0.2692$ & $-0.5363$ & $8.1250$ & $-0.0926$ & $-0.2362$ & $9.4911$ & $-0.2555$ & $-0.5131$ & $8.2307$\\
$1.6$ & $-0.3054$ & $-0.6107$ & $8.1745$ & $-0.1470$ & $-0.3226$ & $9.3597$ & $-0.2934$ & $-0.5888$ & $8.2647$\\
\hline
\end{tabular}
\end{center}
\end{table*}

\begin{table*}[htb]
\begin{center}
\caption{The coefficients $c_{L}$, $c_{H}$, and $c$ for $e^{+}e^{-} \to J/\psi+\chi_{c1}$ at $\sqrt{s}=10.6$ GeV. The unit of $m_c$ is GeV.}
\label{tab: coefxc1}
\begin{tabular}{|c|ccc|ccc|ccc|cc}
\hline
$~$ & \multicolumn{3}{c|}{$A$} & \multicolumn{3}{c|}{$B$} & \multicolumn{3}{c|}{$\sigma$}\\ \hline
$m_c$ & $c_{L}$ & $c_{H}$ & $c$ & $c_{L}$ & $c_{H}$ & $c$ & $c_{L}$ & $c_{H}$ & $c$\\
$1.4$ & $-0.2879$ & $-0.5536$ & $0.3057$ & $-0.2900$ & $-0.5569$ & $2.4524$ & $-0.2883$ & $-0.5543$ & $0.7284$\\
$1.5$ & $-0.3319$ & $-0.6428$ & $0.0986$ & $-0.3348$ & $-0.6477$ & $2.5580$ & $-0.3324$ & $-0.6437$ & $0.5627$\\
$1.6$ & $-0.3725$ & $-0.7327$ & $-0.1510$ & $-0.3764$ & $-0.7400$ & $2.6589$ & $-0.3732$ & $-0.7340$ & $0.3542$\\
\hline
\end{tabular}
\end{center}
\end{table*}

\begin{table*}[htb]
\begin{center}
\caption{The coefficients $c_{L}$, $c_{H}$, and $c$ for $e^{+}e^{-} \to J/\psi+\chi_{c2}$ at $\sqrt{s}=10.6$ GeV. The unit of $m_c$ is GeV.}
\label{tab: coefxc2}
\begin{tabular}{|c|ccc|ccc|ccc|cc}
\hline
$~$ & \multicolumn{3}{c|}{$A$} & \multicolumn{3}{c|}{$B$} & \multicolumn{3}{c|}{$\sigma$}\\ \hline
$m_c$ & $c_{L}$ & $c_{H}$ & $c$ & $c_{L}$ & $c_{H}$ & $c$ & $c_{L}$ & $c_{H}$ & $c$\\
$1.4$ & $-0.4231$ & $-0.7688$ & $-1.3679$ & $-0.5734$ & $-1.0078$ & $-7.0145$ & $-0.4100$ & $-0.7479$ & $-0.8758$\\
$1.5$ & $-0.4648$ & $-0.8687$ & $-0.8550$ & $-0.6462$ & $-1.1770$ & $-7.0557$ & $-0.4520$ & $-0.8470$ & $-0.4205$\\
$1.6$ & $-0.5031$ & $-0.9705$ & $-0.3806$ & $-0.7232$ & $-1.3709$ & $-7.1058$ & $-0.4907$ & $-0.9478$ & $-0.0006$\\
\hline
\end{tabular}
\end{center}
\end{table*}

In our calculations, we use \texttt{FeynArts} \cite{Hahn:2000kx} to generate all the involved Feynman diagrams and the corresponding analytical amplitudes. Then the package \texttt{FeynCalc} \cite{Mertig:1990an} is applied to tackle the traces of the $\gamma$ and color matrixes such that the hard scattering amplitudes are transformed into expressions with loop integrals. In the next step, we utilize our self-written $\textit{Mathematica}$ codes with the implementations of \texttt{Apart} \cite{Feng:2012iq} and \texttt{FIRE} \cite{Smirnov:2008iw} to reduce these loop integrals to a set of irreducible Master Integrals, which could be numerically evaluated by using the package \texttt{LoopTools} \cite{Hahn:1998yk}. 

To check the correctness of our calculations, we simultaneously apply another independent package, Feynman Diagram Calculation (\texttt{FDC}) \cite{Wang:2004du}, to perform the QCD corrections and acquire the same numerical results. As another crucial cross check, with the employment of our numerical expressions, one can immediately reproduce the same values of the $K(=\sigma^{\textrm{NLO}}/\sigma^{\textrm{LO}})$ factors as in refs. \cite{Zhang:2005cha,Gong:2007db,Wang:2011qg,Dong:2011fb}.

\section{Phenomenological results}\label{results}

In order to do the numerical calculations, we choose $m_c=1.5 \pm 0.1$ GeV, $M_{J/\psi}=M_{\eta_c}=M_{\chi_{cJ}}=2m_c$, $\alpha=1/137$, and employ the two-loop $\alpha_s$ running coupling constant. The center-of-mass collision energy is $\sqrt{s}=10.6$ GeV. By using the NRQCD-based results of $\Gamma_{J/\psi \to e^{+}e^{-}}$ and $\Gamma_{\chi_{c2} \to \gamma \gamma}$ of NLO in $\alpha_s$ to respectively match the latest measurements \cite{PDG}, we obtain $|R_{1S}(0)|^2=0.941~\textrm{GeV}^3$ and $|R^{'}_{1P}(0)|^2=0.067~\textrm{GeV}^5$.

\begin{table*}[htb]
\begin{center}
\caption{Total and differential cross sections at $\sqrt{s}=10.6$ GeV. $A$ and $\alpha_{\theta}$ are the coefficients in $\frac{d\sigma}{d\cos\theta}=A(1+{\alpha_\theta}\cos^2\theta)$. $m_c=1.5$ GeV and $\mu_r=2m_c$. }
\label{tab: 2mc}
\begin{tabular}{|c|cccccc|cccccc}
\hline
$~$ & $A_{\textrm{LO}}$(fb) & $A_{\textrm{NLO}}$(fb) & $\alpha_{\theta,\textrm{LO}}$ & $\alpha_{\theta,\textrm{NLO}}$ & $\sigma_{\textrm{LO}}$(fb) & $\sigma_{\textrm{NLO}}$(fb) \\ \hline
$J/\psi+\eta_c$ & $2.744$ & $5.415$ & $1.000$ & $1.000$ & $7.323$ & $14.44$\\
$J/\psi+\chi_{c0}$ & $3.168$ & $4.937$ & $0.252$ & $0.281$ & $6.875$ & $10.81$\\
$J/\psi+\chi_{c1}$ & $0.469$ & $0.410$ & $0.698$ & $0.859$ & $1.158$ & $1.055$\\
$J/\psi+\chi_{c2}$ & $0.893$ & $0.664$ & $-0.198$ & $-0.044$ & $1.670$ & $1.309$\\
\hline
\end{tabular}
\end{center}
\end{table*}

\begin{table*}[htb]
\begin{center}
\caption{Total and differential cross sections at $\sqrt{s}=10.6$ GeV. $A$ and $\alpha_{\theta}$ are the coefficients in $\frac{d\sigma}{d\cos\theta}=A(1+{\alpha_\theta}\cos^2\theta)$. $m_c=1.5$ GeV and $\mu_r=\sqrt{s}/2$. }
\label{tab: halfsqrts}
\begin{tabular}{|c|cccccc|cccccc}
\hline
$~$ & $A_{\textrm{LO}}$(fb) & $A_{\textrm{NLO}}$(fb) & $\alpha_{\theta,\textrm{LO}}$ & $\alpha_{\theta,\textrm{NLO}}$ & $\sigma_{\textrm{LO}}$(fb) & $\sigma_{\textrm{NLO}}$(fb) \\ \hline
$J/\psi+\eta_c$ & $1.856$ & $3.936$ & $1.000$ & $1.000$ & $4.952$ & $10.50$\\
$J/\psi+\chi_{c0}$ & $2.142$ & $3.813$ & $0.252$ & $0.273$ & $4.648$ & $8.328$\\
$J/\psi+\chi_{c1}$ & $0.317$ & $0.386$ & $0.698$ & $0.793$ & $0.783$ & $0.977$\\
$J/\psi+\chi_{c2}$ & $0.604$ & $0.670$ & $-0.198$ & $-0.113$ & $1.129$ & $1.291$\\
\hline
\end{tabular}
\end{center}
\end{table*}

\begin{figure}[!h]
\begin{center}
\hspace{0cm}\includegraphics[width=0.46\textwidth]{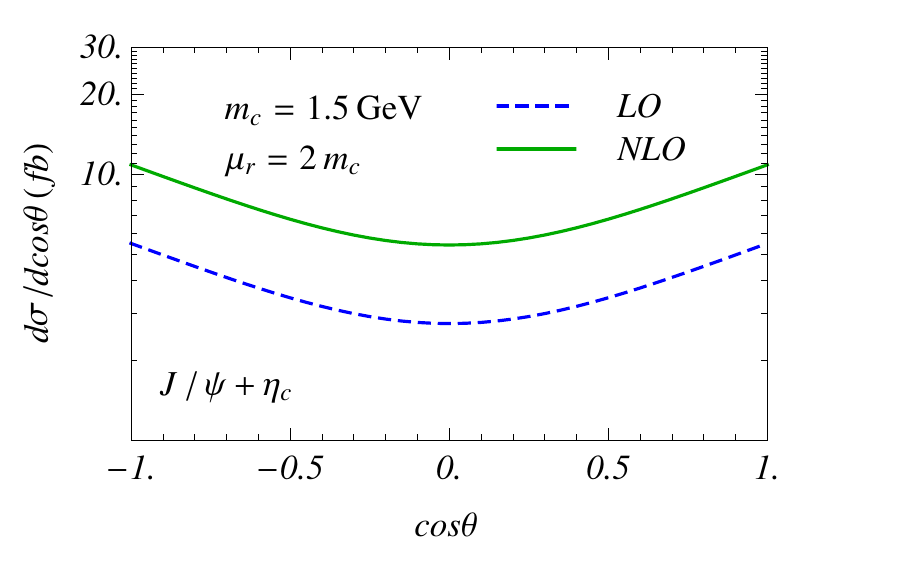}
\hspace{0cm}\includegraphics[width=0.46\textwidth]{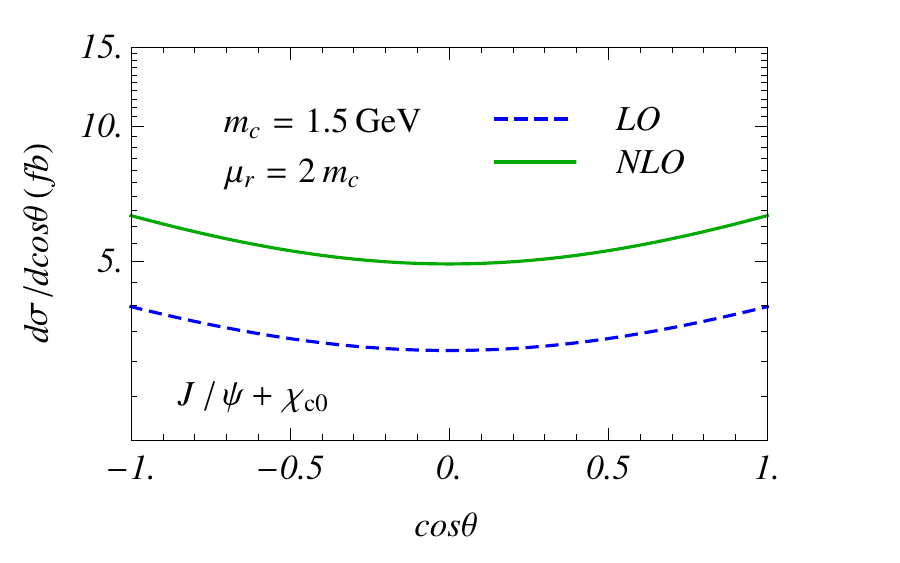}\\
\hspace{0cm}\includegraphics[width=0.46\textwidth]{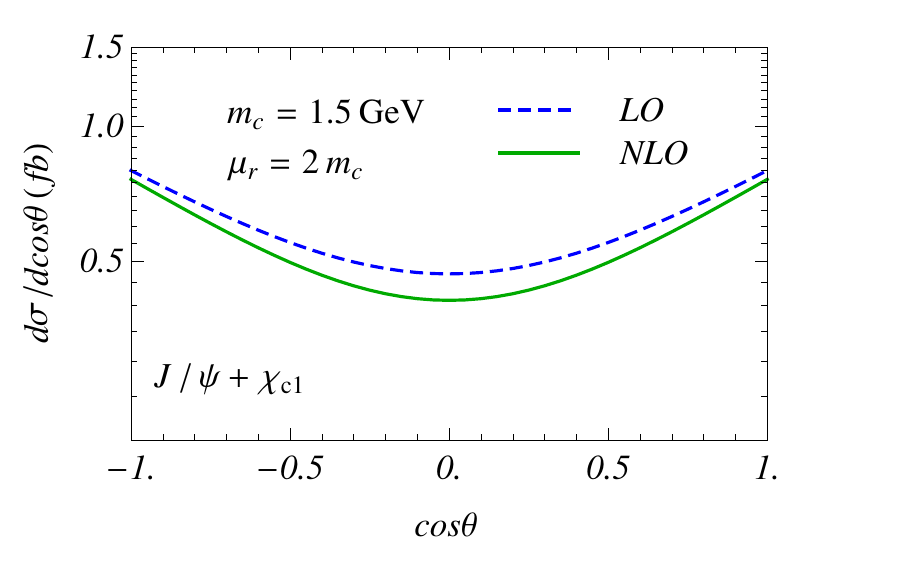}
\hspace{0cm}\includegraphics[width=0.46\textwidth]{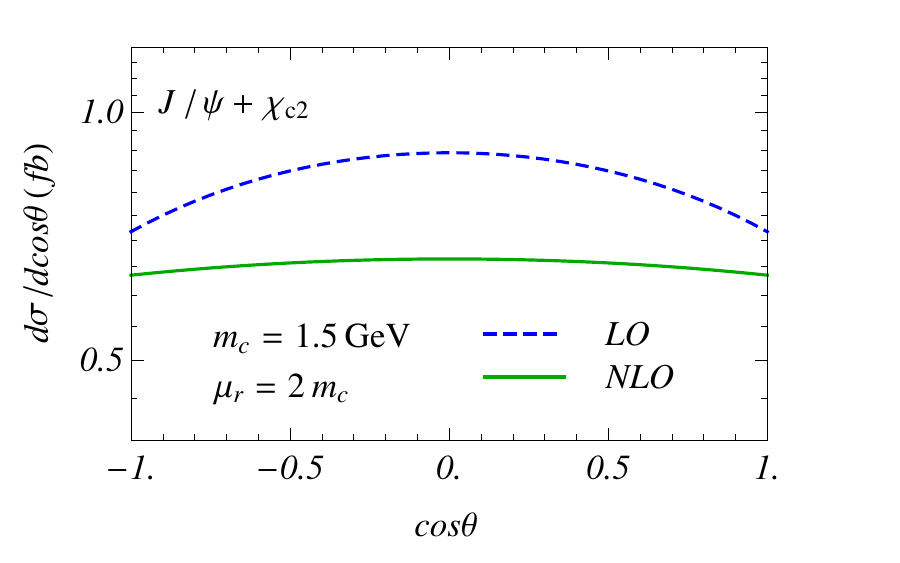}
\caption{\label{fig:dif cross section 1}
$J/\psi$ differential cross sections as a function of $\cos\theta$ at $\sqrt{s}=10.6$ GeV. $m_c=1.5$ GeV and $\mu_r=2m_c$.}
\end{center}
\end{figure}

\begin{figure}[!h]
\begin{center}
\hspace{0cm}\includegraphics[width=0.46\textwidth]{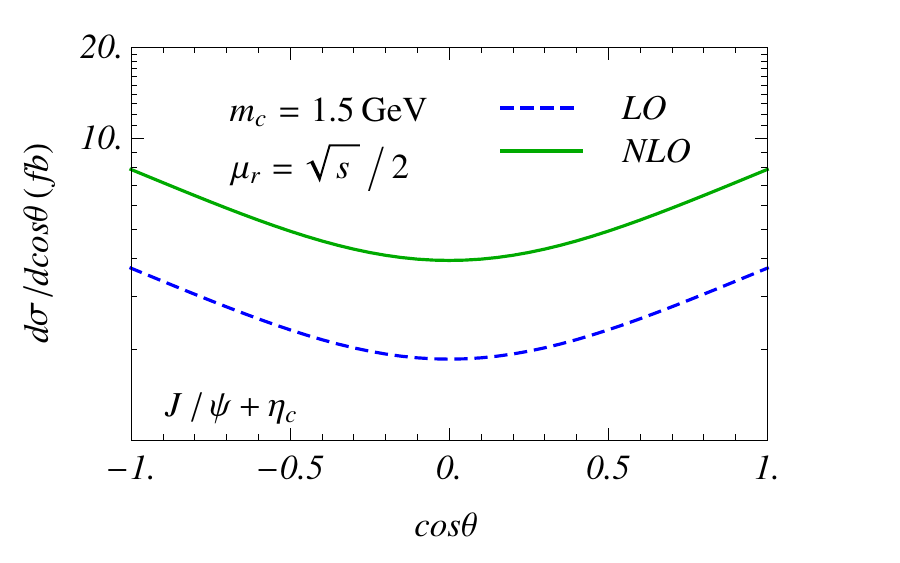}
\hspace{0cm}\includegraphics[width=0.46\textwidth]{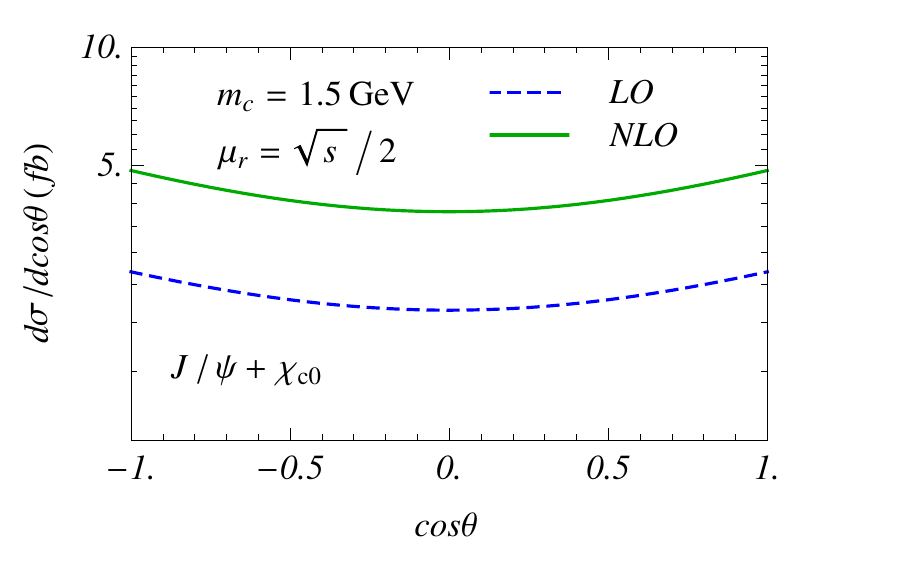}\\
\hspace{0cm}\includegraphics[width=0.46\textwidth]{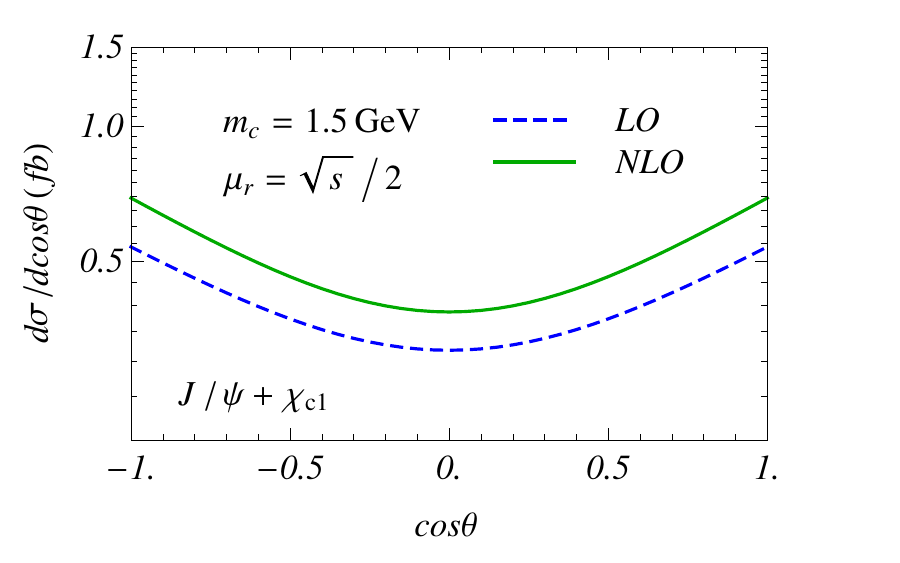}
\hspace{0cm}\includegraphics[width=0.46\textwidth]{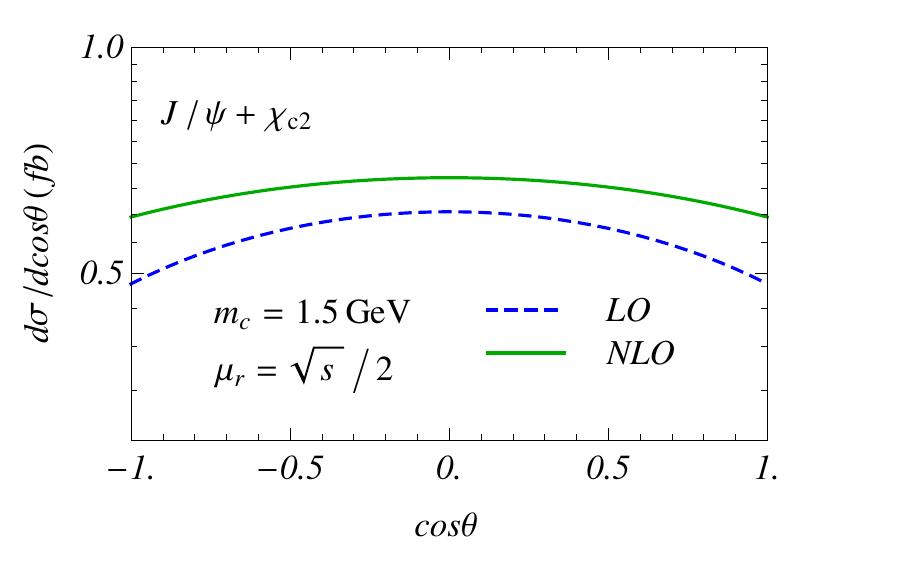}
\caption{\label{fig:dif cross section 2}
$J/\psi$ differential cross sections as a function of $\cos\theta$  at $\sqrt{s}=10.6$ GeV. $m_c=1.5$ GeV and $\mu_r=\sqrt{s}/2$.}
\end{center}
\end{figure}

The NRQCD predictions for the total and differential cross sections are summarized in tables \ref{tab: 2mc} and \ref{tab: halfsqrts}. Inspecting the two tables, on can observe the NLO corrections are crucial for the total cross sections of $e^{+}e^{-} \to J/\psi+\eta_c,\chi_{c0}$, while moderate in the case of $J/\psi$ plus $\chi_{c1,2}$. This can be understood by analyzing the NLO expressions in equation (\ref{NLOexp}). Taking $m_c=1.5$ GeV for instance, the coefficient of $c$ in $\sigma^{\textrm{NLO}}_{\eta_{c}(\chi_{c0})}$, i.e., 12.728(8.2307), would bring forth significant enhancements to the LO results; however, as to $J/\psi+\chi_{c1}(\chi_{c2})$, this coefficient is only 0.5627(-0.4205).

From the data in tables \ref{tab: 2mc} and \ref{tab: halfsqrts}, it is apparent that the LO predictions of ${\alpha_\theta}$ are independent on the choice of the renormalization scale $\mu_r$. With the inclusion of the QCD corrections, ${\alpha_\theta}_{J/\psi+\eta_c}$ remains unchanged; however, by varying $\mu_r$ in $[2m_c, \sqrt{s}/2]$, ${\alpha_\theta}_{J/\psi+\chi_{c0}}$ and ${\alpha_\theta}_{J/\psi+\chi_{c1}}$ are enhanced by about $9-12\%$ and $14-23\%$, respectively, and ${\alpha_\theta}_{J/\psi+\chi_{c2}}$ is reduced in magnitude by about $43-80\%$. These enhancing or reducing effects on ${\alpha_\theta}$ are also clearly visualized by figures \ref{fig:dif cross section 1} and \ref{fig:dif cross section 2}. 

It is worth noting that the impacts of the QCD corrections on ${\alpha_\theta}$ differ substantially from that in the case of total cross section. From equation (\ref{NLOexp}) and tables \ref{tab: coefetac}, \ref{tab: coefxc0}, \ref{tab: coefxc1}, and \ref{tab: coefxc2}, one can see the NLO corrections exert the same (similar) influences in magnitude on $A$ and $B$ in the production of $J/\psi+\eta_c(\chi_{c0,1})$; therefore, ${\alpha_\theta}^{\textrm{NLO}}_{J/\psi+\eta_c}$ is steadily identical to ${\alpha_\theta}^{\textrm{LO}}_{J/\psi+\eta_c}$, and ${\alpha_\theta}^{\textrm{NLO}}_{J/\psi+\chi_{c0,1}}$ resemble their LO results. As for $J/\psi+\chi_{c2}$, the higher-order terms in $\alpha_s$ contributing to $A$ and $B$ are far different from each other, subsequently resulting in the remarkable dissimilarities between ${\alpha_\theta}^{NLO}$ and ${\alpha_\theta}^{LO}$.

\begin{figure}[!h]
\begin{center}
\hspace{0cm}\includegraphics[width=0.46\textwidth]{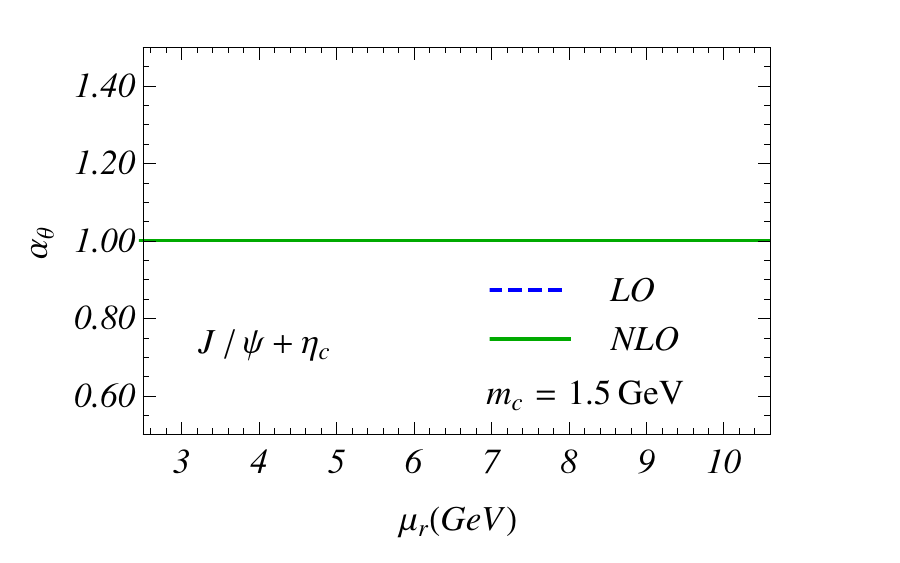}
\hspace{0cm}\includegraphics[width=0.46\textwidth]{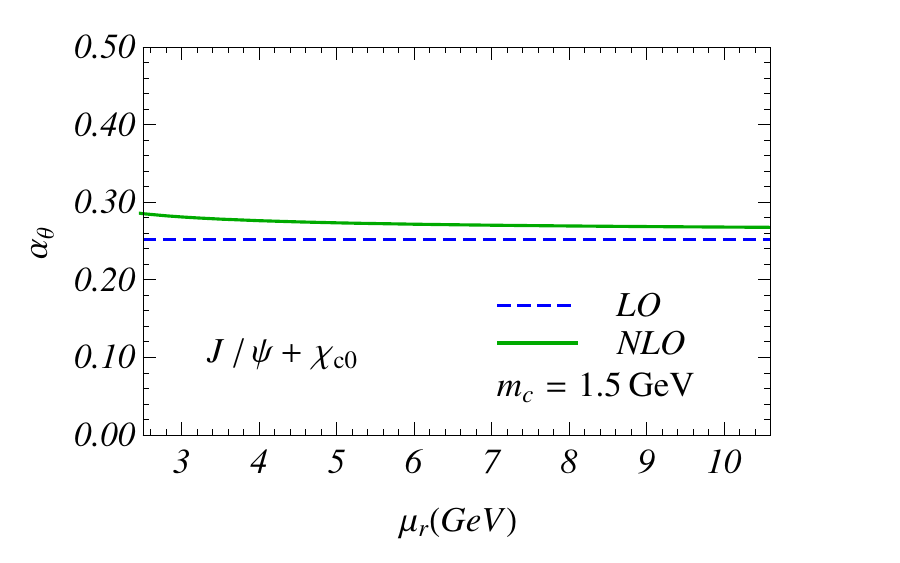}\\
\hspace{0cm}\includegraphics[width=0.46\textwidth]{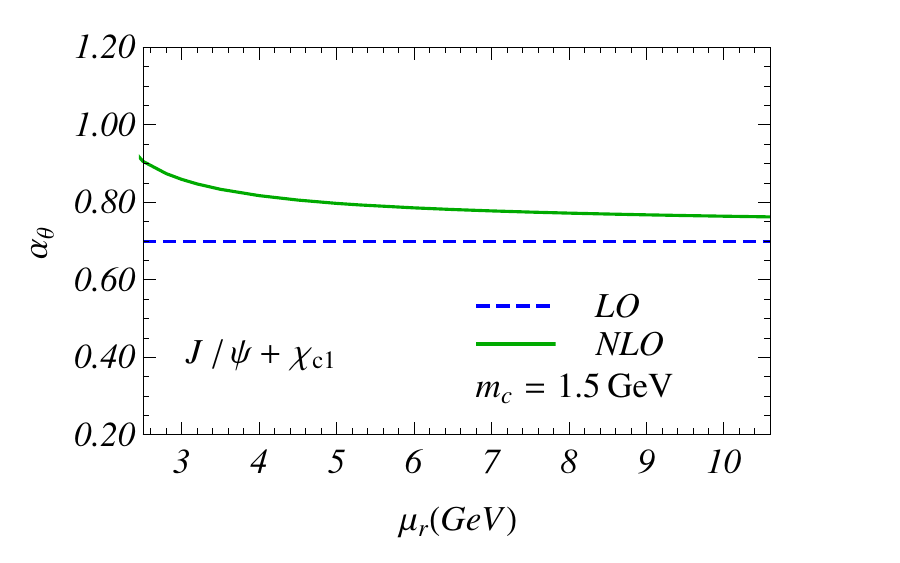}
\hspace{0cm}\includegraphics[width=0.46\textwidth]{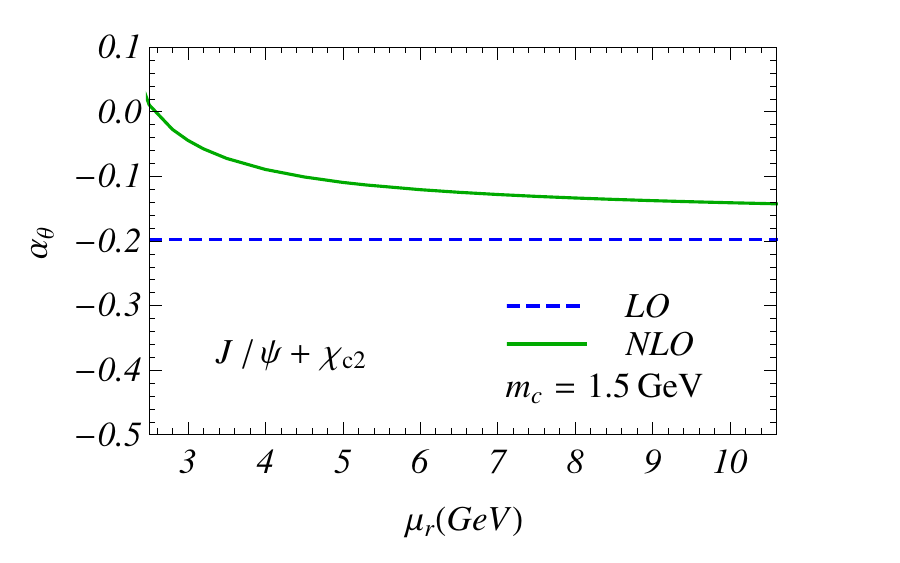}
\caption{\label{fig:alpha}
$\alpha_{\theta}$ parameters as a function of $\mu_r$. $\sqrt{s}=10.6$ GeV and $m_c=1.5$ GeV.}
\end{center}
\end{figure}

In figure \ref{fig:alpha}, we plot the dependence of the predicted ${\alpha_\theta}$ on the renormalization scale, from which one could learn ${\alpha_\theta}^{\textrm{NLO}}_{J/\psi+\eta_c}$ is always identical to 1, and ${\alpha_\theta}^{\textrm{NLO}}_{J/\psi+\chi_{c0}}$ appears to be not sensitive to $\mu_r$ variation; as for $J/\psi+\chi_{c1,2}$, with $\mu_r$ being relatively small, their $\alpha_\theta$ parameters possess a strong $\mu_r$ dependence, which, however, tends to be increasingly mild towards higher $\mu_r$.

\begin{table*}[htb]
\begin{center}
\caption{Comparisons of our predictions of total cross sections (in unit: fb) and $\alpha_{\theta}$ with $\textrm{B}\scriptsize{\textrm{ELLE}}$ and $\textrm{B}\scriptsize{\textrm{ABAR}}$ measurements at $\sqrt{s}=10.6$ GeV. The uncertainties in the predictions are caused by varying $m_c$ in $[1.4,1.6]$ GeV around the central value of $1.5$ GeV. $\mathcal{B}_{>2}$ denotes the branching ratio of $\eta_c(\chi_{cJ})$ into two or more charged tracks.}
\label{tab: exp}
\begin{tabular}{cccccc}
\hline
$~$ & $\textrm{B}\scriptsize{\textrm{ELLE}}(\sigma \times \mathcal{B}_{>2})$ & $\textrm{B}\scriptsize{\textrm{ABAR}}(\sigma \times \mathcal{B}_{>2})$ & $\textrm{NLO}_{\mu_r=2m_c}$ & $\textrm{NLO}_{\mu_r=\sqrt{s}/2}$\\ \hline
$\sigma_{J/\psi+\eta_c}$ & $25.6 \pm 2.8 \pm 3.4$ & $17.6 \pm 2.8 ^{+1.5}_{-2.1}$ & $14.45 ^{+2.835}_{-2.413}$ & $10.50 ^{+1.385}_{-1.320}$\\
$\sigma_{J/\psi+\chi_{c0}}$ & $6.4 \pm 1.7 \pm 1.0$ & $10.3 \pm 2.5 ^{+1.4}_{-1.8}$ & $10.81^{+3.446}_{-2.526}$ & $8.328^{+2.169}_{-1.688}$\\
$\sigma_{J/\psi+\chi_{c1}}$ & - & - & $1.055^{+0.426}_{-0.306}$ & $0.977^{+0.346}_{-0.262}$\\
$\sigma_{J/\psi+\chi_{c2}}$ & - & - & $1.309^{+0.534}_{-0.384}$ & $1.291^{+0.525}_{-0.376}$\\
$\sigma_{J/\psi+\chi_{c1}}$ & \multirow{2}{*}{$<5.3$} & $\multirow{2}{*}{-}$ & $\multirow{2}{*}{$2.364^{+0.960}_{-0.690}$}$ & $\multirow{2}{*}{$2.268^{+0.871}_{-0.638}$}$\\
$+\sigma_{J/\psi+\chi_{c2}}$ & $~$ & $~$ & $~$ & $~$\\
\hline
${\alpha_{\theta}}_{J/\psi+\eta_c}$ & $0.93^{+0.57}_{-0.47}$ & - & $1$ & $1$\\
${\alpha_{\theta}}_{J/\psi+\chi_{c0}}$ & $-1.01^{+0.38}_{-0.33}$ & - & $0.281^{+0.005}_{-0.009}$ & $0.272^{+0.010}_{-0.007}$\\
${\alpha_{\theta}}_{J/\psi+\chi_{c1}}$ & - & - & $0.859^{+0.024}_{-0.025}$ & $0.793^{+0.026}_{-0.027}$\\
${\alpha_{\theta}}_{J/\psi+\chi_{c2}}$ & - & - & $-0.044^{+0.014}_{-0.015}$ & $-0.113^{+0.030}_{-0.036}$\\
\hline
\end{tabular}
\end{center}
\end{table*}

At last, we confront our predictions with experiment. As can be seen in table \ref{tab: exp}, the calculated $\sigma_{J/\psi+\eta_c}$ is consistent, within uncertainties, with the $\textrm{B}\scriptsize{\textrm{ABAR}}$ result, but still inferior to the central value of $\textrm{B}\scriptsize{\textrm{ELLE}}$ measurement; $\sigma_{J/\psi+\chi_{c0}}$ agrees with the data within errors. The sum of $\sigma_{J/\psi+\chi_{c1}}$ and $\sigma_{J/\psi+\chi_{c2}}$ is compatible with the upper limit given by the $\textrm{B}\scriptsize{\textrm{ELLE}}$ Collaboration. From the uncertainties induced by varying $m_c$ in $[1.4,1.6]$ GeV around 1.5 GeV, it is inferred that the $c$-quark mass ambiguities significantly\footnote{Under the same extracting strategy as in ref. \cite{Wang:2011qg}, by which the extracted $|R_{1S}(0)|^2$ and $|R^{'}_{1P}(0)|^2$ depend on charm-quark mass $m_c$, one would reproduce the slight uncertainties therein caused by the variation of $m_c$.}, or even primarily, affect the calculated total cross sections.

With regards to the $\alpha_{\theta}$ parameter, we find the predicted ${\alpha_{\theta}}_{J/\psi+\eta_c}$ agrees well with the $\textrm{B}\scriptsize{\textrm{ELLE}}$ result; however, the increase of $\alpha_{\theta}$ in $J/\psi+\chi_{c0}$ stemming from the incorporation of the QCD corrections further intensifies the disturbing discrepancy in existence between theory and experiment. ${\alpha_{\theta}}_{J/\psi+\chi_{c0}}$ and ${\alpha_{\theta}}_{J/\psi+\chi_{c1}}$ exhibit strong stability under the $m_c$ variation, while the deviation in $m_c$ from 1.5 GeV by $\pm 0.1$ GeV would bring about a $30\%$ fluctuation of ${\alpha_{\theta}}_{J/\psi+\chi_{c2}}$.

In consideration of the noticeable disagreement between the theoretical result of ${\alpha_{\theta}}_{J/\psi+\chi_{c0}}$ and $\textrm{B}\scriptsize{\textrm{ELLE}}$ data, which can hardly be remedied by properly choosing the values of $m_c$, $\mu_r$, and NRQCD LDME, it seems to be premature to draw a decisive conclusion concerning the experimental verification of the NRQCD description of $e^{+}e^{-} \to J/\psi+\chi_{c0}$ from the coincidence between total cross section and experiment. To shed light on this issue, it would be desirable to extend the existing measurements to include $e^{+}e^{-} \to J/\psi+\chi_{c1}$ and $e^{+}e^{-} \to J/\psi+\chi_{c2}$, especially at the Super-$B$ factories which are designed to run with a luminosity up to $\sim 10^{36}\textrm{cm}^{-2}\textrm{s}^{-1}$.

\section{Summary}\label{sum}

In order to provide deeper insight into the well-known exclusive double-charmonium productions in $e^{+}e^{-}$ annihilation at $B$ factories, we in this paper carry out the first NLO study of the $J/\psi$ angular distributions in $e^{+}e^{-} \to J/\psi+\eta_c,\chi_{cJ}$ with $J=0,1,2$ based on the NRQCD factorization. The numerical results show that the newly-calculated QCD corrections moderately affect the LO predictions of ${\alpha_{\theta}}_{J/\psi+\chi_{c0}}$ and ${\alpha_{\theta}}_{J/\psi+\chi_{c1}}$, while significantly reduce in magnitude the LO result of ${\alpha_{\theta}}_{J/\psi+\chi_{c2}}$. Concerning the production of $J/\psi+\eta_c$, the QCD corrections do not change its $\alpha_\theta$ value. By comparisons to experiment, we find ${\alpha_{\theta}}_{J/\psi+\eta_c}$ is consistent with the $\textrm{B}\scriptsize{\textrm{ELLE}}$ measurement; however, the existing radical incompatibility between the LO prediction of ${\alpha_{\theta}}_{J/\psi+\chi_{c0}}$ and experiment becomes even worse by including the QCD corrections.

\acknowledgments
This work is supported in part by the Natural Science Foundation of China under the Grant No. 11705034 and No. 12065006, and by the Project of GuiZhou Provincial Department of Science and Technology under Grant No. QKHJC[2019]1160 and No. QKHJC[2020]1Y035. \\

\providecommand{\href}[2]{#2}\begingroup\raggedright

\end{document}